\documentclass[aps,prl,reprint,superscriptaddress,floatfix,longbibliography]{revtex4-2}

\usepackage{amsmath,amssymb,bm}
\usepackage{graphicx}
\usepackage[colorlinks=true,linkcolor=blue,citecolor=blue,urlcolor=blue]{hyperref}
\usepackage[mathlines]{lineno}
\graphicspath{{figures/}{../figures/}}

\begin{document}
\title{Fifth-Harmonic Feedback Controls the Existence of a Third-Harmonic Zero}

\author{Kanchan Sarkar}
\email{pcksarkar@gmail.com}
\affiliation{Institut f\"ur Theoretische Chemie, Universit\"at Ulm, Oberberghof 7, 89081 Ulm, Germany}
\date{\today}

\begin{abstract}
An exact zero of a generated harmonic needs two real cancellations; a damped oscillator's leading fundamental-source balance supplies one. Absorber loss leaves a phase deficit of fixed sign, amplitude independent at leading order, closed instead by the orbit's own $5\Omega$ motion returning to $3\Omega$ at weight $|\Delta_1|^4$. Closure is therefore unreachable at weak amplitude: the channel supplying it ranks at $O(F^7)$, not at the leading $O(F^3)$. The fifth harmonic is $1.8\times10^{-3}$ of the relative-coordinate fundamental, yet a balance homotopy weighting it below $0.48$ annihilates the tracked pair in a saddle-node.
\end{abstract}

\maketitle

A transmission zero suppresses one response component while motion persists elsewhere. Two interfering pathways cancel exactly at one real frequency, as in the configuration-interaction zero of a Fano lineshape~\cite{Fano1961,Miroshnichenko2010} or the antiresonance of a conservative vibration absorber~\cite{DenHartog1928,denHartog1985,ewins2009modal}. The cancellation is fragile. In the damped two-mode system considered here, absorber loss carries the transfer-function zero off the real axis, and a real-frequency sweep finds only a finite notch. Nonlinear absorbers and antiresonances are usually tracked through such response extrema~\cite{RenaultThomasMahe2019,DetrouxEtAl2015,ShamiGiraudAudineThomas2022}, and nonlinear resonators separately exhibit amplitude-dependent frequency shifts, modal interactions, and harmonic generation~\cite{AldridgeCleland2005,WestraEtAl2010PRL,DykmanEtAl2019PRL,HouriEtAl2020PRL,BachtoldMoserDykman2022,SamantaEtAl2023,SommerEtAl2025PRL,LiEtAl2026PRL}.

Under single-tone excitation, a generated harmonic has no direct forcing term; its complex response is built entirely from nonlinear sources. A minimum on a one-parameter frequency sweep imposes one real stationarity condition, whereas an exact zero requires both response quadratures to vanish and generically needs two real controls. That distinction is operational: at a full-rank zero, two controls can re-null both quadratures after small parameter drift; a finite notch retains a complex residual. In the present system the orbit itself supplies an additional complex source direction through its generated harmonics. We ask whether that feedback can support a zero branch that the fundamental-only balance does not contain.

Cancellation between nonlinear generation pathways, backaction from generated fields, and engineered nonlinear nulls all have precedents~\cite{DoronEtAl2019,LiEtAl2018THG,BertiEtAl2019,SarkarRay2019,kayal2025higher,KenigEtAl2012}, reviewed against the present result in the End Matter. The zero reported below is absent from the nearby fundamental-only balance and appears only when the self-consistent $5\Omega\to3\Omega$ return is retained, whereas a second zero of the same equations is already present in the $\{1,3\}$ weak-drive balance.

\begin{figure*}[t]
    \centering
    \includegraphics[width=\textwidth]{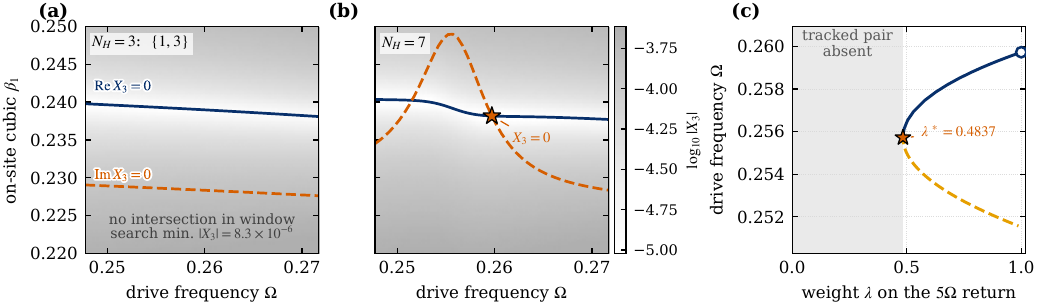}
    \caption{Retaining the fifth harmonic changes whether the third-harmonic zero exists. (a)~At $N_H=3$, so that only the odd content $\{1,3\}$ is kept, the contours $\mathrm{Re}\,X_3=0$ (solid) and $\mathrm{Im}\,X_3=0$ (dashed) run without meeting anywhere in the window; the smallest sampled $|X_3|$ on this grid is $8.3\times10^{-6}$. The panel is a detail of the wider $\Omega\in[0.12,0.45]$, $\beta_1\in[0.05,0.60]$ scan reported in the text, whose minimum is $4.1\times10^{-6}$. (b)~At $N_H=7$, on the same axes and the same $\log_{10}|X_3|$ scale, the $\mathrm{Re}$ contour is almost unmoved while the $\mathrm{Im}$ contour is bent up through it, so the two cross transversely at $(\Omega^\ast,\beta_1^\ast)=(0.259728,0.238154)$: the return acts on the quadrature the leading balance cannot close. The crossing is already present at $N_H=5$, the first truncation admitting $5\Omega$; the seventh-harmonic contribution to this balance is below $10^{-9}$, so the change between the two panels is the fifth harmonic and not the higher retained content. (c)~Weighting by $\lambda$ every fifth-indexed cubic contribution to the two $3\Omega$ balance rows, the two roots present at $\lambda=1$ meet in a saddle-node at $\lambda^\ast=0.4837$ and neither continues below it. Weakening the channel removes the tracked pair rather than shifting its location or depth. Panels (a) and (b) are bounded searches over the displayed window.}
    \label{fig:existence}
\end{figure*}

At a fixed harmonic, each nonlinear term acts as a complex source. To compare those sources at the measured coordinate, each must first be propagated through the linear response at that harmonic. If $\mathbf{S}_j$ is a source at $k\Omega$, $\mathbf{D}_k^{-1}$ is the corresponding linear response, and $\mathbf{c}$ selects the measured coordinate, then
\begin{equation}
Y_k=\mathbf{c}^{\dagger}\mathbf{D}_k(\Omega)^{-1}\!\sum_j\mathbf{S}_j.\label{eq:projected}
\end{equation}
An exact zero corresponds to closure of these propagated source vectors in the complex plane.

At finite amplitude, the sources $\mathbf{S}_j$ depend on the periodic orbit itself, so Eq.~\eqref{eq:projected} is self-consistent. We call two contributions distinct pathways only when they arise from separate constitutive terms. Consider a two-degree-of-freedom mechanical realization: a driven mass $x_1$ coupled to an absorber $x_2$:
\begin{align}
\ddot x_1+2\zeta_1\dot x_1+\omega_1^2 x_1+\beta_1 x_1^3&\nonumber\\
{}+\kappa(x_1{-}x_2)+\kappa_{\rm nl}(x_1{-}x_2)^3 &= F\cos\Omega t,\label{eq:model1}\\
\ddot x_2+2\zeta_2\dot x_2+\omega_2^2 x_2&\nonumber\\
{}+\kappa(x_2{-}x_1)+\kappa_{\rm nl}(x_2{-}x_1)^3 &= 0.\label{eq:model2}
\end{align}
All quantities are nondimensional. Unless stated otherwise, $(\omega_1,\omega_2,\zeta_1,\zeta_2)=(1,\,1.25,\,0.015,\,0.02)$, $\kappa=0.10$, and $\kappa_{\rm nl}=-0.30$. Here $\zeta_j$ is a damping rate: the term $2\zeta_j\dot x_j$ gives $Q_j=\omega_j/(2\zeta_j)$ for an uncoupled oscillator.

We write $X_{j,n}$ for the complex $n$th-harmonic coefficient of $x_j$, $X_3\equiv X_{1,3}$ for the measured third harmonic, and $A_n\equiv|X_{1,n}|$ for driven-mass harmonic amplitudes. The full two-index notation is retained whenever the coordinate matters. In the regular weak-drive limit $X_3=O(F^3)$. A generated-harmonic zero means $X_3=0$ while the fundamental motion remains nonzero.

Keeping only the fundamental components in the two cubic source terms gives the leading cancellation condition. Eliminating the absorber from the propagated $3\Omega$ balances yields
\begin{equation}
Z_{22}(3\Omega)\,\beta_1 X_{1,1}^3 + \kappa_{\rm nl}\,L_2(3\Omega)\,(X_{1,1}{-}X_{2,1})^3 = 0,\label{eq:LOzero}
\end{equation}
after cancelling the factor $1/4$ common to both cubic Fourier coefficients, where $L_2(3\Omega)=\omega_2^2-9\Omega^2+6i\zeta_2\Omega$ and $Z_{22}(3\Omega)=L_2(3\Omega)+\kappa$. Absorber damping makes Eq.~\eqref{eq:LOzero} complex, so one real coefficient cannot generically satisfy both quadratures at a prescribed frequency. At the reported frequency, the fundamental-only source reduction would require a complex closing coefficient with $\mathrm{Re}\,\beta_1=0.23112$ and $|\mathrm{Im}\,\beta_1|=9.6\times10^{-4}$, so $\arg\beta_1^{(1)}=4.15\times10^{-3}$ rad. Its real part lies within $2.95\%$ of the converged value, but no real $\beta_1$ closes both quadratures. Across $\Omega\in[0.24,0.28]$ that phase rises monotonically from $3.60\times10^{-3}$ to $5.52\times10^{-3}$ rad without crossing zero. Since $\kappa_{\rm nl}$ enters only as a real factor, no real closing coefficient exists anywhere on that interval. The absence of a nearby leading-order root therefore follows from the phase condition itself. This distinguishes the branch from a detached or isolated resonance curve. Detached curves are well characterized in this class of system, including in the two-cubic absorber model used here, where singularity theory organizes them around a cusp~\cite{CirilloEtAl2017,HabibCirilloKerschen2018,NoelEtAl2015,DetrouxEtAl2018}. Those are statements about the connectivity of a solution family. The exclusion here is instead a sign-definite phase obstruction over an interval, and holds whatever the connectivity turns out to be.

The coupling cubic supplies a second propagated source with a different complex direction. The two null contours then cross transversely in the $(\Omega,\beta_1)$ plane [Fig.~\ref{fig:existence}(b)], where at $N_H=3$ they do not meet at all [Fig.~\ref{fig:existence}(a)]. The root belongs to a one-dimensional locus in $(\Omega,F,\beta_1)$; at the reported point the experimentally natural pair $(\Omega,F)$ has a nonsingular control Jacobian.

Harmonic balance resolves a root at $F=0.30$, $\kappa_{\rm nl}=-0.30$, and $(\Omega^\ast,\beta_1^\ast)=(0.25973,0.23815)$, with $A_1=0.29$. The driven mass is dark at $3\Omega$, but the absorber is not: it retains $|X_{2,3}|=1.6\times10^{-3}$. That surviving motion generates $5\Omega$ in the relative coordinate through $3\Omega+\Omega+\Omega$, and the resulting fifth harmonic mixes back through $5\Omega-\Omega-\Omega$ into the driven-mass $3\Omega$ balance. Because both legs of that loop are cubic, the returned term carries four extra powers of the relative-coordinate fundamental $\Delta_1$ [Eq.~\eqref{eq:em-eta5}]. In a regular weak-drive expansion the leading $3\Omega$ source is $O(F^3)$, so this return is the $O(F^7)$ term: it is not absent from the expansion, but it first appears four more powers of amplitude beyond the balance that fixes the ordinary root. The ordering ranks sources at fixed small amplitude; it is not a description of the branch, which does not reach that limit. What it must cancel is amplitude independent: absorber loss sets the leading phase obstruction, of which the $3\Omega$ feedback removes about a quarter and the $5\Omega$ return the rest. Neither survives $|\Delta_1|\to0$, while the obstruction does. At fixed finite damping, and away from a leading-order phase crossing, closure therefore requires finite response amplitude.

Continuous-time shooting reproduces the same root to relative differences below $10^{-8}$ in both controls, with a nonsingular Jacobian, and returns it for either control pair~\cite{SupplementalMaterial}. The largest Floquet multiplier has modulus $0.69$, so the orbit is locally asymptotically stable.

Spectral truncation corroborates the phase obstruction and exposes the same return path. No root was found for $N_H=3$ or $4$ in a $67\times67$ scan over $\Omega\in[0.12,0.45]$, $\beta_1\in[0.05,0.60]$ refined from $1156$ seeds, whose minimum was $|X_3|=4.1\times10^{-6}$; the branch is first resolved when $5\Omega$ is retained and converges from $N_H=7$. That is bounded-search evidence and does not establish absence outside the window. On the converged orbit, suppressing the fifth and higher harmonics reopens a driven-mass residual $(\mathrm{Re},\mathrm{Im})=(-6.5,6.0)\times10^{-5}$; the $5\Omega-\Omega-\Omega$ mixing term supplies $(+6.5,-6.0)\times10^{-5}$ and closes it to the harmonic-balance tolerance, and the shooting solve then places the corresponding zero directly in the continuous-time ODE~\cite{SupplementalMaterial}.

\begin{figure*}[t]
    \centering
    \includegraphics[width=\textwidth]{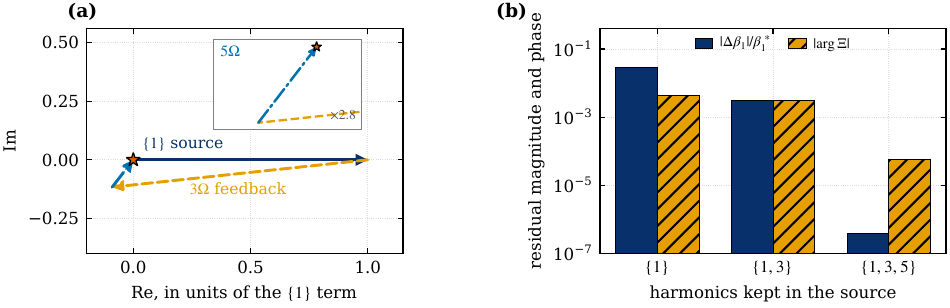}
    \caption{How the closure is apportioned. (a)~Projected third-harmonic source balance, normalized to the fundamental-only term: the $3\Omega$ feedback removes most of the magnitude deficit, and the smaller $5\Omega$ term supplies the remaining complex direction (inset, magnified $2.8\times$). (b)~Error in the closing cubic coefficient, and the phase the complex coefficient would need, for successive source truncations $\{1\}$, $\{1,3\}$, $\{1,3,5\}$.}
    \label{fig:mechanism}
\end{figure*}

The projected balance makes the division of labor visible [Fig.~\ref{fig:mechanism}(a,b)]. With only the fundamental source retained, the complex closing coefficient misses the converged real value by $2.95\%$ and has phase $\arg\beta_1^{(1)}=4.15\times10^{-3}$ rad. Adding the surviving $3\Omega$ motion removes most of the magnitude deficit; including $5\Omega$ closes both to the solver floor, four orders below. The decomposition diagnoses the converged balance; a homotopy shows the return is required. Weighting by $\lambda$ every fifth-indexed cubic contribution to the two $3\Omega$ balance rows, of which the $5\Omega{-}\Omega{-}\Omega$ channel carries $99\%$, and continuing the zero, the locus turns at $\lambda^\ast=0.4837$ and never reaches $\lambda=0$ [Fig.~\ref{fig:existence}(c)]: within this modified harmonic-balance family, reducing the return below $\lambda^\ast$ annihilates the tracked pair in a saddle-node~\cite{SupplementalMaterial}. The partner is a second zero of the unmodified equations, $8.1\times10^{-3}$ away in $\Omega$, so the return is required by both zeros in this window and not by one fragile root.

The same equations also contain a perturbative third-harmonic root. It is already present at $N_H=3$, satisfies the leading-order condition, and tends to $(\Omega,\beta_1)\to(1.2485,-0.0249)$ with $A_1\propto F$ as $F\to0$~\cite{SupplementalMaterial}. The low-frequency branch instead lies between two turning points of the drive on one connected locus~\cite{SupplementalMaterial}. Toward lower frequency it turns at $F=0.2577$, where $A_1$ takes its minimum of $0.2508$, and the drive then rises again. Toward higher frequency the same locus carries a minimum of the drive at $F=0.02099$, a factor of fourteen below the working point, reached near the lower coupled mode with $A_1/F=32$; $A_1$ never falls below $0.2508$ anywhere on the traced locus. The drive reaches a minimum there and rises again, to $F=2.41$. That minimum is set by dissipation. Averaging each equation against its own velocity over a period gives the exact balance $F\langle\cos\Omega t\,\dot x_1\rangle=2\zeta_1\langle\dot x_1^2\rangle+2\zeta_2\langle\dot x_2^2\rangle$, so every periodic orbit obeys $F\ge2\zeta_1\Omega A_1$. Along this stretch the amplitude is locked to $0.4\%$, at the value $|\Delta_1|=(-4\kappa/3\kappa_{\rm nl})^{1/2}$ where the effective fundamental coupling $\kappa+\tfrac34\kappa_{\rm nl}|\Delta_1|^2$ cancels; the absorber then decouples at $\Omega$ while staying lit at $3\Omega$, and the driven mass rides its own Duffing backbone. The fold sits at that backbone, where the bound is nearly tight: $2\zeta_1\Omega A_1=0.020982$ against a computed fold at $0.020989$, and the bound holds at every coupling tracked~\cite{SupplementalMaterial}. Two roots of one set of equations therefore behave in opposite ways under the same reduction of the drive.

At the baseline point, the small $5\Omega$ correction is strongly amplified by the spectrum. The absorber tuning $\omega_2/\omega_1=5/4$ gives coupled modes $(\omega_-,\omega_+)=(1.0406,1.2961)$, and $5\Omega^\ast=1.2986$ lies only $0.20\%$ from $\omega_+$. The relative-coordinate gain at $5\Omega$ accordingly exceeds that at $3\Omega$ by $9.8$. The fifth harmonic is nevertheless small, because its source is: it is generated through the surviving absorber third harmonic. It carries $1.8\times10^{-3}$ of the relative-coordinate fundamental, the channel the coupling cubic drives and reads, and $2.9\times10^{-4}$ of the driven-mass fundamental. It is the term that closes the residual quadrature.

That amplification is a property of the working point, not of the whole locus. At the resonant end, near $\Omega\simeq1.05$, $|g_5|$ falls by a factor of $337$ while the response amplitude rises and the orbit becomes strongly nonlinear; the zero is still there, but the small-rotation expansion behind Eq.~\eqref{eq:em-sufficient} no longer applies and we have not resolved the closure into channels there~\cite{SupplementalMaterial}. The channel-resolved statements below are established at the working point and its near-linear neighborhood.

Near the transverse root a coefficient error produces a proportional response, a $20$~dB loss of suppression per decade; the drive pair tracks the zero across the operating window, and the zero is re-nulled, with the control pair appropriate to each case, under absorber detuning, mass-ratio, quintic and quadratic perturbations and beyond the coupling fold~\cite{SupplementalMaterial}. Two limits are reached within the tested ranges: drive-only access is lost between $\zeta_2=0.03$ and $0.04$, and at coupling quintic $k_5=0.15$ a root is recovered at $F=0.34$ but not at $F=0.30$~\cite{SupplementalMaterial}.

The same zero remains Floquet stable under common loss reduction over four decades, to $Q\sim3\times10^5$, with the working point converging~\cite{SupplementalMaterial}. Over that continuation the absolute detuning of $5\Omega^\ast$ from $\omega_+$ grows only from $0.20\%$ to $0.83\%$ while the relative-coordinate gain rises from $25.9$ to $48.7$~\cite{SupplementalMaterial}.

Hardening and softening on-site nonlinearities are available in gated and strain-engineered resonators~\cite{Kozinsky2006,LiStrain2024}; the less direct ingredient is the softening cubic in the relative coordinate. The polynomial coupling used above is only a local model because $\kappa_{\rm nl}<0$ makes its quartic potential unbounded at large displacement. Its tangent stiffness remains positive over the reported operating interval, and a positive quintic regularization preserves the zero~\cite{SupplementalMaterial}. More directly, the Supplemental Material replaces the polynomial coupling by the exact odd force of a symmetric electrostatic coupler. The zero remains Floquet stable and the drive--bias control map remains full rank both at the baseline and at the top of the common-damping continuation~\cite{SupplementalMaterial}. The operating point is still close to static pull-in, so this is a realizability check rather than a device design.

One further constraint follows from the working point itself. Since $\Omega^\ast\simeq0.26\,\omega_1$, the driven mass responds quasi-statically, with $A_1\omega_1^2/F=0.97$, and the force needed to reach the operating amplitude is set by stiffness rather than by $Q$. Raising the quality factor removes the phase obstruction without easing the drive requirement. The low-forcing resonant segment of the same locus, near $\omega_-$ with $A_1/F=32$, is where that requirement relaxes~\cite{SupplementalMaterial}.

A zero of a generated harmonic can therefore be carried by a term that is fourth order in the response amplitude, invisible to any balance truncated at the fundamental and ranking in a regular weak-drive expansion only at $O(F^7)$. That is what separates this branch from the ordinary root of the same equations: it stays finite in amplitude over the traced locus, and at the working point weighting its closing channel down destroys it. Shooting locates the same zero in the continuous-time flow, and it survives common-loss continuation and replacement of the polynomial coupling by an exact electrostatic force law, with a full-rank control map that lets two parameters re-null both quadratures after drift.

The homotopy is a deformation of the balance equations rather than of a physical system, so its discriminating power has to be tested where the operating point, not the residual, is what changes. A second oscillator model with inertial coupling supplies that test~\cite{SupplementalMaterial}. There the closure estimate, with one prefactor carried over from the mechanical model, identifies the spectral region where such a zero should sit; solving in that region returns a zero showing the defining signatures of the mechanical case: onset at $N_H=5$, the same ordering of source truncations, and a homotopy fold at $\lambda^\ast=0.2925$. A second zero of those same equations, at a frequency where $5\Omega$ lies $20\%$ from the upper mode, shows none of them: the leading-order coefficient is wrong by $83\%$, the truncation hierarchy does not converge, and its homotopy passes through $\lambda=0$ without a fold. The signatures therefore follow spectral placement.

The same cubic mixing architecture suggests an analogous $m\to m{+}2\to m$ loop at other odd harmonics; we have tested it at $m=3$ only. The fifth harmonic carries only $1.8\times10^{-3}$ of the relative-coordinate fundamental, yet it decides whether the third-harmonic zero exists. Low-order harmonic balance is known to generate spurious branches; the failure here runs the other way. The retained spectral weight is unchanged to a part in $10^3$, and the harmonic that gets discarded is the one that fixes existence.

\begin{acknowledgments}

I thank Professor Axel Gro\ss{} (Universit\"at Ulm) and Professor Pranab Sarkar (Visva-Bharati University) for their guidance and for many valuable discussions. Every derivation was checked and every reported number regenerated by the author from the deposited scripts, and no figure image was produced or altered by a generative model. The free version of ChatGPT was used for language editing, and Anthropic Claude for language editing and for drafting analysis and figure-rendering scripts; Sec.~S15 of the Supplemental Material records the scope of that use in full. I acknowledge support from the state of Baden-W\"urttemberg through bwHPC and from the German Research Foundation (DFG) through grant INST 40/575-1 FUGG (bwForCluster JUSTUS~2), as well as from the Dr.~Barbara Mez-Starck Foundation.

\end{acknowledgments}

\textit{Data availability}---The code and data supporting this article are
openly available in an archived deposit under a persistent
DOI~\cite{SarkarDataZenodo}. The deposit contains a
reproduction guide mapping each quantitative claim and figure to the script and
data used to generate it.

\nocite{ChiconeODE2006,HairerNorsettWanner1993,KrackGross2019,Seydel2010}
\bibliography{refs_v2}

\makeatletter
\if@twocolumn\onecolumngrid\fi
\makeatother
\begin{center}
\textbf{\large End Matter}
\end{center}
\vspace{0.4em}
\makeatletter
\if@twocolumn\twocolumngrid\fi
\makeatother
\setcounter{equation}{0}
\renewcommand{\theequation}{EM\arabic{equation}}

\textit{Relation to prior mechanisms.}---Interference-based third-harmonic control has been treated in both effective descriptions and fully coupled nonlinear mode equations. Kayal and Ray reported higher-harmonic antiresonance within a perturbative coupled-oscillator treatment~\cite{kayal2025higher}. Equation~\eqref{eq:LOzero} is the corresponding leading-order condition for the present model. It follows from eliminating the absorber across both propagated $3\Omega$ balances, so the generated harmonic of the second oscillator is retained and the response at the generated frequency carries the coupled determinant $(L_1(3\Omega)+\kappa)Z_{22}(3\Omega)-\kappa^2$ of Eq.~\eqref{eq:em-onepath} rather than a single-oscillator denominator. Its root at $(1.2485,-0.0249)$ as $F\to0$ is this system's analogue of the perturbative antiresonance, and is the branch used for comparison throughout. Vibrational antiresonance in nonlinear coupled oscillators has been treated in the same framework~\cite{SarkarRay2019}, and antiresonances in nonlinear structures are routinely continued in two parameters~\cite{RenaultThomasMahe2019,DetrouxEtAl2015}. Equations~\eqref{eq:model1}--\eqref{eq:model2} are themselves the standard two-cubic nonlinear absorber, whose resonance structure, detached branches, and quintic regularization have been analyzed by singularity theory~\cite{CirilloEtAl2017,HabibCirilloKerschen2018}; the object here is a zero of a generated harmonic rather than a response extremum. Nonlinear cancellation has also been engineered deliberately in nanomechanical resonators~\cite{KenigEtAl2012}. Li \textit{et al.} include strong-pump self-consistency in a microcavity model~\cite{LiEtAl2018THG}. Doron, Michaeli, and Ellenbogen obtain direct--cascaded cancellation through an effective hyperpolarizability~\cite{DoronEtAl2019}. Earlier atomic-physics work found coherent cancellation involving a third-harmonic field generated in the same medium~\cite{PayneGarrettFerrell1986}. Feedback from self-generated harmonics is also known in nonlinear optical propagation~\cite{BertiEtAl2019}.

The setting is standard; the branch is not. The low-frequency branch is a root of the self-consistent periodic orbit, and the term that closes its balance enters at $O(|\Delta_1|^4)$, so no leading-order reduction produces it. Eq.~\eqref{eq:LOzero} has no root near it, and the branch stays finite-amplitude along the traced continuation. The comparison is against a second root of the same equations, computed with the same solver and truncation family~\cite{SupplementalMaterial}.

Harmonic cancellation, generated-field backaction, and harmonic feedback each have precedents. What has not been reported is a zero excluded from the fundamental-only balance by a sign-definite phase obstruction and resolved only once the orbit's own $5\Omega$ motion is fed back, in a system that also carries an ordinary $\{1,3\}$ root. Unless a symmetry or structural relation locks the two quadratures, a complex zero has codimension two.

\textit{The two leading-order reductions.}---We use two related approximations. The \emph{fundamental-only balance}, Eq.~\eqref{eq:LOzero}, evaluates the cubic sources using the fundamental of the actual orbit and reproduces $\beta_1^\ast$ to $3.0\%$. The \emph{linear-fundamental reduction}, Eq.~\eqref{eq:em-scalefree}, also replaces the fundamental amplitude ratio by its linear response and is about $6\%$ low. In this paragraph write $X_j\equiv X_{j,1}$. Retaining only the fundamental, $x_j\simeq\mathrm{Re}(X_j e^{i\Omega t})$, a cubic generates a third-harmonic coefficient proportional to $X_j^3/4$. The on-site and coupling cubics drive the $3\Omega$ balance through $\beta_1X_1^3/4$ and $\kappa_{\rm nl}(X_1-X_2)^3/4$, and eliminating the absorber amplitude returns Eq.~\eqref{eq:LOzero}: the two propagated cubic sources must cancel as complex numbers. The linear fundamental response gives $X_2/X_1=\kappa/Z_{22}(\Omega)$. Dividing by $X_1^3$ removes the overall amplitude and yields
\begin{equation}
Z_{22}(3\Omega)\,\beta_1+\kappa_{\rm nl}\,L_2(3\Omega)\Big(1-\tfrac{\kappa}{Z_{22}(\Omega)}\Big)^{3}=0,
\label{eq:em-scalefree}
\end{equation}
within this leading-order reduction. In this reduction the result is independent of forcing and primary damping. Absorber damping enters through $L_2$ and $Z_{22}$, so the equation remains complex: cancellation at fixed $\Omega$ would need the two propagated source directions to be real-linearly dependent, and on the branch reported here absorber damping prevents that alignment.

\textit{Phase reduction and a closure criterion.}---Solving the projected balance for the on-site coefficient reduces the zero to a phase condition. Wherever $\mathcal{A}_3[x_1^3]\neq0$,
\begin{equation}
\beta_1=\Xi,\quad
\Xi=-\frac{\kappa_{\rm nl}L_2(3\Omega)\,\mathcal{A}_3[d^3]}{Z_{22}(3\Omega)\,\mathcal{A}_3[x_1^3]},
\label{eq:em-xi}
\end{equation}
where the two Fourier coefficients are evaluated on the orbit. Thus $\Xi=\Xi(\Omega,\beta_1;F)$, and Eq.~\eqref{eq:em-xi} is an implicit fixed-point condition rather than a formula for $\beta_1$. Write $H_1=\mathrm{Re}\,\Xi-\beta_1$ and $H_2=\mathrm{Im}\,\Xi$. Where $\partial H_1/\partial\beta_1\neq0$, the condition $H_1=0$ defines $\beta_1=b(\Omega)$ locally, and the zero reduces to a root of the real function $h(\Omega)=H_2[\Omega,b(\Omega)]$. A sign change of $h$ is the phase criterion.

At leading order the orbit dependence drops out and $\Xi\to\Xi_{\rm LO}(\Omega)$, Eq.~\eqref{eq:em-scalefree}, so the required coefficient depends on frequency but not on amplitude or $\beta_1$. A real $\beta_1$ requires phase $0$ modulo $\pi$. The Letter gives the monotone phase obstruction across $\Omega\in[0.24,0.28]$. A scan of $\mathrm{Im}\,\Xi_{\rm LO}$ over $\Omega\in(0.02,3.0)$, which contains the whole traced locus, finds only two crossings: the separate perturbative branch near $\Omega\simeq\omega_2$, and a near-divergence of $|\Xi_{\rm LO}|$ inside the upper-mode resonance. Both require $\beta_1<0$, and the nearest stays $0.99$ away in $\Omega$ at every absorber loss tested~\cite{SupplementalMaterial}. That bounds the weak-drive behavior of the branch: any family of exact zeros with $F\to0$, $\Omega\to\Omega_0\in(0,3)$, $\beta_1$ bounded and $X_{1,1}/F\to\chi_1(\Omega_0)\neq0$ must satisfy $\beta_1\to\Xi_{\rm LO}(\Omega_0)$ with $\mathrm{Im}\,\Xi_{\rm LO}(\Omega_0)=0$, since the higher channels are then relatively $O(F^2)$ and $O(F^4)$, whereas $\beta_1$ stays within $[0.235,0.299]$ at every traced point. The zero-frequency endpoint is separate, because $\mathrm{Im}\,\Xi_{\rm LO}$ vanishes there identically; it is closed by the slope, $\mathrm{Im}\,\Xi_{\rm LO}=c_0\Omega+O(\Omega^3)$ with $c_0=2.16\times10^{-3}>0$, so no crossing occurs at any small $\Omega>0$. Reaching either endpoint would require a sign-changing continuation through $\beta_1=0$, not observed on the traced component; global connectivity of the zero set is not established, nor is the joint corner $(F,\Omega)\to(0,0)$ or a limit with $\beta_1\to\infty$~\cite{SupplementalMaterial}.

At the fixed baseline frequency $\Omega_0^\ast=0.259728$, the leading-order phase is proportional to absorber loss, with $\arg\Xi_{\rm LO}(\Omega_0^\ast)/\zeta_2=0.220$ as $\zeta_2$ is reduced over four decades. That path holds $\zeta_1$ fixed and moves $\Omega^\ast$. Along the common-damping continuation, where the working point converges, the same loss scaling is observed: $\arg\Xi_{\rm LO}/\epsilon$ and $\mathrm{Im}\,\eta_5/\epsilon$ remain constant to three digits while the magnitude correction tends to $2.85\%$. Higher harmonics rotate the phase, $\Xi=\Xi_{\rm LO}(1+\eta_3+\eta_5+\cdots)$, and the fifth-harmonic term scales as
\begin{equation}
\eta_5\ \sim\ \frac{\kappa_{\rm nl}^{2}\,g_5(\Omega)}{Z_{22}(3\Omega)}\,|\Delta_1|^{4},\qquad
g_5=(1,-1)\mathbf{D}_5^{-1}(1,-1)^{\!\top},
\label{eq:em-eta5}
\end{equation}
with $\Delta_1$ the relative-coordinate fundamental and $g_5$ the complex relative-coordinate propagator, whose magnitude is the gain reported in the Supplemental Material. At leading order this feedback term scales as $|\Delta_1|^4$. In a regular weak-drive expansion $\Delta_1=O(F)$, so the source hierarchy at $3\Omega$ reads $S_3=F^3S_3^{(3)}+F^5S_3^{(5)}+F^7S_3^{(7)}+\cdots$ and the $3\Omega\!\to\!5\Omega\!\to\!3\Omega$ return sits in the $F^7$ term, four powers of amplitude beyond the $F^3$ balance that fixes the ordinary root. The deficit it must cancel, $\arg\Xi_{\rm LO}$, is proportional to absorber loss and independent of amplitude, so closure requires $|\Delta_1|$ above a loss-set floor rather than being reachable as $F\to0$. Over the finite interval used in the numerical check the fitted effective exponent is $3.59$, with the $|\Delta_1|^4$ law recovered only asymptotically, and no weak-amplitude limit is reached on the traced branch. The propagator $g_5$ grows large where $5\Omega\simeq\omega_+$, which is what makes the correction efficient at the baseline point.

On an interval where $\mathcal{A}_3[x_1^3]\neq0$ and $\partial H_1/\partial\beta_1\neq0$, so that $b(\Omega)$ exists and $h$ is continuous, a sign change of $h$ gives a root by the intermediate-value theorem; equivalently, the two quadrature-null contours of Fig.~\ref{fig:existence}(b) cross. The reduction estimates when the feedback is large enough to produce that sign change,
\begin{equation}
|\mathrm{Im}\,\eta_5(\Omega)|\ \gtrsim\ |\arg\Xi_{\rm LO}(\Omega)+\mathrm{Im}\,\eta_3(\Omega)|.
\label{eq:em-sufficient}
\end{equation}
The third-harmonic channel rotates the phase first, so the fifth harmonic must cancel the residual $\arg\Xi_{\rm LO}+\mathrm{Im}\,\eta_3$, not the leading phase alone. On the converged orbit these contributions are $4.395\times10^{-3}$ and $-1.193\times10^{-3}$, against $\mathrm{Im}\,\eta_5=-3.259\times10^{-3}$. On the locus these numbers cannot test Eq.~\eqref{eq:em-sufficient}, since the zero condition forces them to balance; they show how the rotation is partitioned, the $3\Omega$ channel taking about a quarter and the $5\Omega$ channel the rest. Equation~\eqref{eq:em-sufficient} is therefore testable only off the locus, and the inequality carries an unevaluated $O(1)$ prefactor. It applies to the near-linear part of the branch, where $|\Delta_1|\propto F$; the forcing at which the arm turns comes from the fold continuation instead. Under common loss scaling both the leading phase deficit and $\mathrm{Im}\,\eta_5$ fall with the loss while the magnitude correction stays finite~\cite{SupplementalMaterial}.

That turn is fixed instead by the fold curve $\kappa^\ast(F)$. For the present $m=3$ branch, closure combines a damping-split leading phase, an amplitude-dependent higher-harmonic feedback channel, and two real controls. The scaling suggests, but does not demonstrate, an analogous role for an $(m{+}2)\Omega$ channel at other odd harmonics.

\textit{One-pathway limit.}---Setting $\kappa_{\rm nl}=0$ while retaining the full periodic waveform gives
\begin{equation}
X_3=\frac{-\beta_1\mathcal{A}_3[x_1^3]\;Z_{22}(3\Omega)}
{\big(L_1(3\Omega)+\kappa\big)Z_{22}(3\Omega)-\kappa^2},
\label{eq:em-onepath}
\end{equation}
where $\mathcal{A}_3$ extracts the third-harmonic Fourier coefficient. The expression agrees with the numerical balance to relative error $2\times10^{-14}$. A zero can arise through $Z_{22}(3\Omega)=0$, which absorber damping excludes at real nonzero frequency, or through $\mathcal{A}_3[x_1^3]=0$, which is algebraically possible because mixing terms within one cubic can cancel; within the stated window $|\mathcal{A}_3[x_1^3]|$ stays bounded away from zero and the one-pathway branch has a finite minimum, and no source zero was found in the sampled range outside it either~\cite{SupplementalMaterial}. At the working point the one-pathway amplitude is $|X_3|=2.8\times10^{-3}$, against an integrator-floor residual of order $10^{-12}$ from independent time integration of the two-pathway orbit.

\textit{Transversality.}---At the working point the control map is nonsingular, with condition number $5.1$ in $(\Omega,\beta_1)$ and $2.5$ in $(\Omega,F)$; the corresponding determinants are $3.3\times10^{-5}$ and $3.1\times10^{-6}$, whose scale depends on the choice of variables. Continuous-time shooting reproduces both from a full-rank $6\times6$ shooting Jacobian~\cite{SupplementalMaterial}. This is a local statement; global uniqueness and global structural stability are not established.

\textit{Finite-amplitude closure and fold.}---The raw and projected balances use different residual coordinates. In the raw driven-mass residual, the fifth harmonic closes the balance. After eliminating the correlated absorber residual, the projected balance assigns most of the magnitude correction to $3\Omega$ feedback and the remaining phase direction to $5\Omega$ [Fig.~\ref{fig:mechanism}]. The decomposition is diagnostic and does not determine the orbit. The forcing threshold at fixed $\kappa$ and the coupling fold at fixed $F$ are two slices of the same fold curve $\kappa^\ast(F)$. It crosses the baseline coupling between $F=0.256$ and $F=0.258$, where continuation at that coupling turns the arm~\cite{SupplementalMaterial}.

\end{document}